\documentclass{article}

\usepackage{graphicx}
\usepackage{amssymb}
\usepackage{epstopdf}
\usepackage{amsmath}
\usepackage{natbib}
\usepackage{subfigure}

\def\pd#1#2{\frac{\partial #1}{\partial #2}}

\begin{document}
\bigskip

{\bf Deep-water gravity waves: nonlinear theory of wave groups}
\bigskip

I. M. Mindlin 

Department of Applied Mathematics, State Technical University,

Minina St., 24, Nizhny Novgorod 603155, Russia 

E-mail address: ilia.mindlin@gmail.com
\bigskip

Nonlinear initial-boundary value problem on deep-water gravity waves 
of finite amplitude is solved approximately (up to small terms 
of higher order) assuming that the waves are generated by 
an initial disturbance to the water and the horizontal dimensions 
of the initially disturbed body of the water are much larger than 
the magnitude of the free surface displacement. 

A numerable set of specific free surface waves is obtained in 
closed form and it is shown that free surface waves produced by 
an arbitrary initial disturbance to the water is a combination 
(not superposition: the waves are nonlinear) of the specific waves. 

A set of dispersive wave packets is found with one-to-one 
correspondence between the packets and positive integers, say, 
packet numbers, such that any initial free surface displacement 
gradually disintegrates into a number (limited or unlimited, 
depending on initial conditions) of the wave packets. 
The greater the packet number, the shorter the wavelength 
of the packet's carrier wave component, the slower the packet 
travels, the slower the packet disperses; 
evolution of any of the packets is not influenced by evolution 
of any other one. 

It is found that in case of infinitesimal wave amplitude the present
theory is in agreement with linear wave theory. On the other hand, the
behaviour of wave packets of large numbers and asymptotic behaviour of
solutions of Schr\"odinger equation for weakly nonlinear waves are found
to be similar, except for dispersion  
The theory is tested against experiments performed in a water tank. 
\bigskip

\bigskip

{\bf 1. Introduction  }
\bigskip

{\bf 1.1. Overview of the suggested approach} 
\bigskip

We present a theory of nonlinear deep-water waves in a vertical plane which start to propagate away from an initially disturbed body of water. Then the water is acted on by no external force other than gravity. It is assumed that the free surface of the water is infinite in extent and the pressure along the surface is constant.

Classical equations of the problem includes
(i) the Laplace's equation for the velocity potential, 
(ii) nonlinear kinematic condition which means that a liquid
particle in the free surface can have no velocity relative to the
surface in the direction of the normal, (iii) nonlinear condition
of the pressure continuity across the free surface,
(iiii) conditions at infinity, (iiiii) initial conditions. 

In the available literature nonlinear analysis of deep water waves is based on
Schr\"odinger equation for weakly nonlinear waves.

Assuming that the frequency, amplitude, and wave number of a weakly nonlinear
 evolving wave train are slowly varying 
functions of time, the classical equations for free surface gravity waves were 
reduced to equations  governing the evolution of the frequency, amplitude, and 
wave number.  These equations in turn were reduced to nonlinear Schr\"odinger 
equation for complex-valued wave envelope [1, 2, 3, 4]. 

Asymptotic behaviour of solutions of the Schr\"odinger equation was revealed 
by Zakharov \& Shabat [5]. Provided the initial free surface displacement (away from 
the equilibrium  position of the surface) decays sufficiently rapidly with 
distance from the maximum of the displacement, it was found, in particular, 
that an initial wave packet of arbitrary envelope shape disintegrates with 
time into a number of final packets, each of specific envelope shape, and 
an oscillating 'tail'; the envelope of each final packet is a solution of 
the Schr\"odinger equation;  the number of final wave packets depends on 
initial conditions for the equation; the envelope of a final packet travels 
at a constant velocity, keeps its shape, and retains its shape and velocity 
after interaction with other final packets (this is the reason to use the term 'envelope soliton'\,). That is, the solitons do not disperse. 

Disintegration of a wave packet into final packets travelling at different 
velocities is also the dispersion. But the final packets don't disperse.  Why? 
Supposedly due to a dynamic balance between nonlinear and dispersive effects. 

In any case, existence of the envelope solitons for solutions of Schr\"odinger equation does not assure the balance of nonlinearity and dispersion for solutions of classical equations for deep-water gravity waves.

A quite different approach to the initial-boundary value problem on the 
waves is developed in the article, based on the author's previous works [6, 7] . 

The principal features of the suggested approach are: 
\begin{itemize}
\item
It is assumed that horizontal dimensions of the initially disturbed body 
of water are much larger than the vertical displacement of the free surface 
in the wave originating area. 
\item
A nonlinear transformation of the equations is performed. 
The transformation maps the trace of the free surface in a plane of flow onto 
a half of a circle. The mapping allows us to use a discrete (with respect to spacial argument) set of functions instead of integral Fourier transform involving continuum of  harmonic waves.
\item
It is also assumed that initial conditions can be expanded in trigonometric series convergent on the segment of the circle.
\item
Solution of the transformed equations is sought in the form of series in powers of small parameter, the ratio of the two linear dimensions of initial disturbance.
\item
Adopting the initial conditions in the form of trigonometric series, we express the free surface evolution in the form of parametric equations, which allow waves to steepen until overturning occurs.
\item
The solution is obtained under boundary conditions at infinity which require the energy supplied to the water by a source of disturbances to be finite. By the conditions, the plane waves should be expressed in terms of functions vanishing at infinity along the free surface. A complete numerable set of such functions is constructed in the article. 
\end{itemize}  

Thus, in the suggested approach, different from replacing the original equations with the Schr\"odinger equation and seeking the 'envelope solution',
\begin{itemize}
\item
no assumptions are made about wave numbers, frequency spectrum or any other solution features.The assumptions concern only initial conditions.
\item
relatively small terms are omitted in the solution, not in the equations, since small terms in the equations may essentially influence the long time evolution of the solutions.
\item
no a-priory assumption about time evolution of the solution is made.
\item
the energy of the solution is always finite. Consequently, periodic waves on the free surface infinite in extent (including sinusoidal and progressive Stokes waves) are 'prohibited' by these conditions. 
\end{itemize}
\bigskip
. 
{\bf 1.2. Brief overview of the resulting solutions} 
\bigskip

Solutions presented in this article describe waves expressed in terms of nonlinear combinations of specific wave packets. The packets are subject to dispersion. 
 A wave group on the free 
surface is a mixture of finite or infinite (it depends on initial conditions) 
set of the specific wave packets, evolution of each packet 
in the mixture is not influenced by evolution of the others.

Though any specific packet disperses, it tends to keep its shape within a finite time interval which depends on the packet carrier wave length. 
The greater the packet's number, the longer the time interval. 
So, on a finite time interval, behaviour of the specific packets of large 
numbers seems to be similar to that of the "envelope soliton" solutions 
of the Schr\"odinger equation [5].

Applicability of the Schr\"odinger equation to deep water waves is substantiated by
some results of experiments performed in a water tank [8].
But theoretical results of the manuscript are supported by the same experiments 
in no less extent. Moreover, the theoretical results of the manuscript are in extremely 
good quantitative agreement with the experiments presented in [9].

As a limit case of the nonlinear theory,  equations for infinitesimal deep water waves are obtained with boundary conditions on the evolving free surface, and the solution to the equations is found in the form of linear superposition of wave packets..  
If the boundary conditions are shifted from the evolving free surface to the equilibrium position of the surface, the equations reduce to equations of classical linear theory of infinitesimal waves.
\bigskip

{\bf 1.3. Article outline} 
\bigskip

This work is organized and presented in eleven sections.
Problem outline, basic equations in curvilinear coordinates and solution to the equations 
(up to small terms of higher order) are presented in sections 2 - 6.
The results of the theory are discussed in sections 7, 8, 9 and 10. Properties of the specific wave packets are described and illustrated in sections 7 and 8. In section 9, a set of infinitesimal linear wave packets are obtained as a limit of the set of nonlinear packets. 
 In connection with the limit, we would like to draw special attention to the  assertions 2 and 3 formulated at the end of sections 9  and 10 respectively. 

In the section 10, the theory is tested against experimental results 
pre\-sen\-ted in Yuen \& Lake [8] and Feir [9].  The test show quantitative agreement between theoretical predictions and the experimental measurements of some parameters of wave packets. 

The possibility to obtain the higher-order approximation to the full nonlinear problem is discussed in section 11.

\bigskip

{\bf 2. Problem outline. }
\bigskip                                                                      

Referring to figure 1 and  assuming that the motion is two-dimensional, 
in the $(x,y)$-plane,
consider flow of an ideal heavy uniform liquid of density 
$\gamma$; the $x$-axis is oriented upward and the $y$-axis in horizontal
direction. 
We think of the liquid as being contained between two vertical planes
(parallel to the $(x,y)$-plane) at a distance apart. 

Let the curve $\,\,\Gamma\,\,$ (in figure 1) be the trace of the free
surface $\,\,S\,\,$ in the $\,\,(x,y)\,\,$ plane,
$\,x=f<0,\,$ $y=0$ be the coordinates of the pole $\,O_1\,$ of the polar
coordinate system in the $\,(x,y)\,$ plane,
$\,\theta\,$ be the polar angle measured from the positive $x$-axis
in the counterclockwise direction, $\,t\,$ be the time. 
The external pressure, $\,P_*\,$, on the free surface is constant.
The liquid fills the space below the free surface.
The equilibrium position of the free surface is in horizontal plane $x=0$. 
\begin{figure}
	\resizebox{\textwidth}{!}
		{\includegraphics{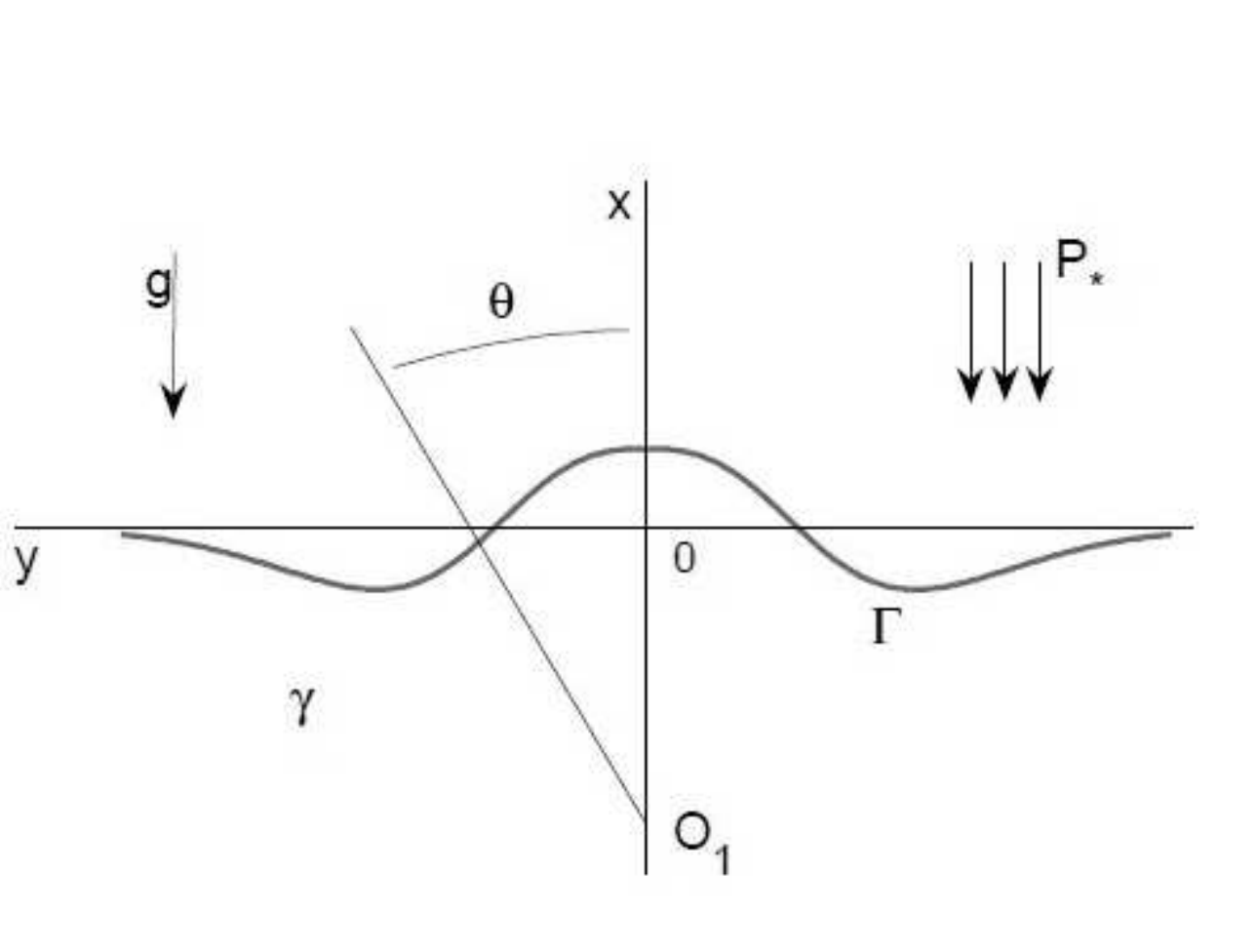}} 
	\caption{
	Coordinate systems and sketch of the free surface of a liquid. 
	}
	\label{qu-.1-.10}
\end{figure} 

To satisfy the Laplace's equation, the velocity potential $\Phi$ is sought 
in the form of a doublet distribution over the free surface: 
$$
\Phi(P)=\frac {1}{2\pi}\int_{\Gamma}\nu(Q)\pd {}{n}
\left(\ln\frac {1}{r}\right)dl 
        $$
where $Q$ and $P$ are two points in the $(x,\,y)$-plane; $Q$ is a point on the curve $\Gamma$, 
point $P$ does not belong to the curve, $r=|PQ|$ is the distance between the points, $\nu(Q)$ is the density of the doublets distributed over the curve $\Gamma$,  $dl$ is the infinitesimal element of the curve (Fig. 1).
Notation $\partial {}/{\partial n} $ is used for directional derivative at the point $Q$ 
in the orthogonal direction to the curve $\Gamma$ (unit vector of the direction points in the region above the free surface). 

Then the boudary-integral equation method is used: the classical differential equations accompanied by boundary conditions are reduced to a system of integrodifferential equations involving functions defined on the evolving free surface unknown in advance [10] .
\bigskip

{\bf 3. Basic equations in curvilinear coordinates }
\bigskip

Equations of the free surface are sought in the parametric form
  $$
 x=W(\theta,t),\, y=(W-f)\,\tan\,\theta,\,    
-\pi/2 < \theta < \pi/2                          \eqno(3.1)  
                     $$
where $\,\,W(\theta,t)\,\,$ is an unknown function that must be found while
solving the problem. 

Formally, equations (3.1) for each specified function
$\,\,W\,\,$ describe a family of curves depending on $\,f,\,$ $\,t\,$
 being considered as constant. The value  of $\,\,f\,\,$ 
determines the horizontal scale of the problem. 

In the $\,\,(x,y)\,\,$ plane, curvilinear coordinates $(\sigma, \theta)$
are defined by the relations
$$
x=\sigma+W(\theta,t),\ \  y=(\sigma+W-f)\,\tan\,\theta
	$$
so the equation of the free surface takes the form $\sigma =0$ (the liquid
occupies the half-space $\sigma <0$). 

All equations are rewritten in the curvilinear coordinates and 
the problem is formulated mathematically in terms of
nonlinear in\-teg\-ro\-dif\-fe\-ren\-ti\-al equations 
in two unknown functions $W(\theta,t)$ and $\nu(\theta,t)$.

Note, that the change of variables is a nonlinear transformation of the equations written  in Cartesian coordinates.

For any function $F(\sigma ,\theta ,t)$ the following notation is introduced for one-sided limit:  
$$F_{-}=\lim_{\sigma\to -0}F(\sigma ,\theta ,t)$$

Velocity potential in curvilinear coordinates $(\sigma, \theta)$ is expressed as
$$
\Phi(\sigma,\theta,t)=\frac{f}{|f|}\frac{1}{2\pi}\int\limits_{-\pi/2}^{\pi/2}
\nu(\theta_1,t)A(\sigma,\sigma_1,\theta,\theta_1,t)
\left.\frac{d\theta_1}{R}\right|_{\sigma_1=0}      \eqno(3.2)   
        $$
$$
A=(\sigma -\sigma_1+W-W_1)(f-W_1)+
        $$
$$
\frac{\sigma-f+W}{\sigma_1-f+W_1}\,\,\pd {W_1}{\theta_1}\,\,
(\tan\theta\cdot\cos^2\theta_1-
\sin\theta_1\cdot\cos\theta_1)
        $$     
$$
R=(\sigma-\sigma_1+W-W_1)^2\cos^2\theta_1+
        $$
$$
[(\sigma-f+W)\tan\theta\cdot\cos\theta_1-
(\sigma_1-f+W_1)\sin\theta_1]^2
        $$
$$
W=W(\theta,t),\,\,\,\,\,\,W_1=W(\theta_1,t),\,\,\,\,\,\,
        $$
Equations (3.3) and (3.4) are nonlinear kinematic condition (ii) and  nonlinear condition 
(iii) of the pressure con\-ti\-nu\-i\-ty across the free surface $\sigma=0$ respectively:
$$
\pd {W}{t}=D_2\,\pd {\Phi}{\sigma_-}-D_1\,\pd {\Phi}{\theta_-}    \eqno(3.3)   
        $$
$$
\pd {\Phi}{t_-}-\frac{1}{2}
D_2\left(\pd {\Phi}{\sigma_-}\right)^2+\frac{1}{2}
\left(D\pd {\Phi}{\theta_-}\right)^2+W(\theta,t)=0    \eqno(3.4)     
        $$
$$
D=\frac{\cos\theta}{W-f},\,\,\,\,\,\,
D_1=\left(\sin\theta+D\,\pd{W}{\theta}\right)D 
	$$
$$
D_2=1+(D_1+D\sin\theta)\pd {W}{\theta}
        $$
The conditions at infinity are taken in the form
$$
|W(\theta ,t)|<C(t)\cos^2\theta,\,\,\,\,\, \lim \limits_{\cos \theta \to 0}
\frac {\partial W}{\partial \theta}=0,\,\,\,\,\,
|\nu(\theta,t)|<C(t)                                    \eqno(3.5)    
        $$
Although the liquid is infinite in extent, these conditions assure that
the total energy supplied to the liquid by a source of disturbances is finite.

All equations are written in non-dimensional variables.
Since the problem has no characteristic linear size, 
the dimensional unit of length, $\,\,L_*,\,\,$ is a free parameter.
But for applications in the section 10, the value of $L_*$,    
as well as the value of $|f|L_*$, will be obtained  from instrumental data.
The dimensional unit of time, $\,\,T_*,\,\,$ is defined
by the relation $\,\,T_*^2g=L_*\,$, where $\,\,g\,\,$ is the acceleration
of free fall. The non-dimensional acceleration of free fall is equal to unity.
All parameters, variables and equations are made non-dimensional by 
the quantities $\,L_*,\,T_*,\,P_* $ and 
the density of water $\gamma_*=1000\,$ $\hbox{kg/m}^3$.
\bigskip

{\bf 4. Leading-order equations}
\bigskip

Let $c$ be the maximum of the free surface displacement 
in the wave origin, so $\,\varepsilon=c/f$ is the ratio
of the maximum displacement to the characteristic horizontal dimension of the origin. 

We seek expansions of the form  
$$
W(\theta,t)=c [W_0(\theta,t)+\varepsilon W_1(\theta,t)+...], \,\,\,\,\,\,\,
\nu(\theta,t)=c\cdot [\nu_0(\theta,t)+\varepsilon\nu_1(\theta,t)+...],\,\,\,\,\,\,
	$$
$$
\Phi(\sigma,\theta,t)=c\cdot [\Phi_0(\sigma,\theta,t)+\varepsilon\Phi_1(\sigma,\theta,t)+...] 
	$$ 
By expanding all terms of the equations (3.2) - (3.4)
in powers of $\varepsilon$ and equating coefficients of like powers of $\varepsilon$ we obtain the leading-order equations in unknown functions  $W_0(\theta,t),\,\,\nu_0(\theta,t),\,\,\,\Phi_0(\sigma,\theta, t)$: 
$$
\pd {\Phi_0}{t_-}+W_0=0  \eqno(4.1)      
        $$
$$
 \pd {W_0}{t}=\pd {\Phi_0}{\sigma_-}+\frac{\sin\theta\cdot\cos\theta}{f}\,      
\pd {\Phi_0}{\theta_-}                                        \eqno(4.2)  
        $$
$$
\Phi_0(\sigma,\theta,t)=\frac{\sigma}{|\sigma|} H(\nu_0)
        $$
where the integral operator $H$ is defined as follows
$$
H(F(\theta))=|\alpha(1-\alpha)|\int\limits_{-\pi/2}^{\pi/2}
\frac{F(\theta_1)\,d\theta_1}{a\cos^2\theta_1+b\sin^2\theta_1-c\sin(2\theta_1)}
	$$
$$
\alpha=\frac{f}{f-\sigma},\,\,\,\,\,\,a=(1-\alpha)^2+\tan^2\theta,\,\,\,\,\,\,
b=\alpha^2,\,\,\,\,\,\,c=\alpha\,\,\tan\theta
	$$
Elemntary technique for analytic evaluation of integrals gives
$$
2H(c_1\cos (2n\theta)+c_2\sin (2n\theta))=
 \mu_n(c_1\cos (2n\beta)+c_2\sin (2n\beta))     \eqno(4.3) 
        $$

$$
\cos (2 \beta)=\frac {2\alpha\cos^2\theta -1}{z},\,\,\,\,\,\,
\sin (2 \beta)=\frac {\alpha \sin (2\theta)}{z}
        $$
$$
\alpha=\frac {f}{f-\sigma },\,\,\,\,\,\,
z=\sqrt {1-4\alpha(1-\alpha)\cos^2\theta}
        $$
$$
\mu_n=\frac {1}{z^n},\,\,\,\,\hbox{if}\,\,\,\, \alpha (1-\alpha)<0
        $$
$$ 
\mu_n=z^n,\,\,\,\,\hbox{if}\,\,\,\, \alpha (1-\alpha)>0
	$$

With the use of (4.3), equations (4.1), (4.2) accompanied by boundary conditions (3.5) and initial conditions
$$
W_0(\theta,0)=\cos^2\theta \sum_{n=1}^{+\infty}
[a_n\cos(2n\theta)+b_n\sin(2n\theta)]                  \eqno(4.4)  
	$$
$$
\nu_0(\theta,0)= \sum_{n=1}^{+\infty}[\rho_n\cos(2n\theta)+e_n\sin(2n\theta)]                    
	$$
can be solved exactly.

The constants $a_n$, $b_n$ determine the initial displacement to the free surface,  
the constants $\rho_n$, $e_n$ determine the initial velocity field. 

The  technique for handling the problem is described in [6].  
The principal steps of the technique are shown in Appendix A. 
\bigskip

{\bf 5. Exact series solution to the leading-order equations}
\bigskip

The exact solution of the problem (4.1), (4.2), (4.4), (3.5) was obtained 
in the form of functional double series [6]: 
$$ 
x=cW_0(\theta,t),\\\ y=(x-f)\tan\theta, \\\
-\pi/2< \theta < \pi/2                                      \eqno(5.1)     
	$$         
$$
W_0(\theta,t)= \sum_{n=1}^{+\infty} (-1)^{n-1}\frac{1}{2n}
[a_nC_n(\tau,\theta)+b_nS_n(\tau,\theta)]+                 
        $$
                        	$$          \eqno(5.2)   $$     
$$
\frac{1}{\sqrt{2|f|}}\sum_{n=1}^{+\infty} (-1)^n
[\rho_nH_n(\tau,\theta)+e_nG_n(\tau,\theta)] 
      $$
$$
\pd{\nu_0}{t}=2W_0(\theta,t),\,\,\,\,\,\,t=\tau\,{\sqrt{2|f|}} 
	$$
where
$$
C_n(\tau,\theta)=-2\cos^2\theta\cdot\sum_{k=1}^{+\infty}
(-1)^kl_{kn}(\tau)\cos(2k\theta)                       \eqno(5.3)     
        $$
$$
S_n(\tau,\theta)=-2\cos^2\theta\cdot\sum_{k=1}^{+\infty}
(-1)^kl_{kn}(\tau)\sin(2k\theta)                       \eqno(5.4)     
        $$
$$
H_n(\tau,\theta)=-2\cos^2\theta\cdot\sum_{k=1}^{+\infty}(-1)^k
h_{kn}(\tau)\cos(2k\theta)                          \eqno(5.5)        
        $$
$$
G_n(\tau,\theta)=-2\cos^2\theta\cdot\sum_{k=1}^{+\infty}(-1)^k
h_{kn}(\tau)\sin(2k\theta)                          \eqno(5.6)        
        $$
$$
l_{kn}(\tau)=2\int\limits_0^{+\infty}
x^3e^{-x^2}L_{k-1}^{(1)}(x^2)L_{n-1}^{(1)}(x^2)\cos(\tau x)\,dx \eqno(5.7)   
        $$
$$
l_{kk}(0)=k,\,\,\,\,\,\,l_{kn}(0)=0,\,\,\,\,\hbox{for}\,\,\,\,k\ne n
        $$
$$
h_{kn}(\tau)=2\int\limits_0^{+\infty}
x^2e^{-x^2}L_{k-1}^{(1)}(x^2)L_{n-1}^{(1)}(x^2)\sin(\tau x)\,dx   
        $$
The right hand side of (5.7) involves the Laguerre polynomials,
$\,\,L_k^{(1)}(u),\,\,$ defined by the recursion relation [11]
$$
 L_0^{(1)}=1,\,\,\,L_1^{(1)}(u)=2-u,\,\,\,\,\hbox{for}\, k\ge 2
	$$
$$
kL_k^{(1)}(u)=(2k-u)L_{k-1}^{(1)}(u)-kL_{k-2}^{(1)}(u)     \eqno(5.8)   
	$$
Though the function $W_0(\theta,t)$ is a linear combination of solutions of linear equations (4.1), (4.2), the waves (5.1) are still nonlinear. 
In the wave theory, the principle of linear superposition 
states, in particular, that when two waves overlap, 
the actual displacement of any point of the free surface, at any time, 
is obtained by adding two displacements. The waves (5.1) does not obey the 
principle (the implicit form of the free surface  
is $x=cW_0(\hbox{arc tangent}(y/(x-f)),t)$). 
\bigskip

{\bf 6. Sum up of the series involved in the leading-order solution } 
\bigskip 

Two theorems are formulated and proved in this section.

{\bf Theorem 1.}   
In any rectangle $|\theta|\le\pi/2-\delta$,
 $ 0\le \tau \le T,\,\,$  \linebreak 
($0<\delta<\pi/2,\, T>0 $
may  be assigned arbitrarily) series (5.3) through (5.6) converge
uniformly with regard to $ \theta $ and $ \tau $.
\bigskip

{\bf Theorem 2.} On the interval $\,|\theta|<\pi/2\,$  the sums of the series 
are given by
$$
C_n(\tau,\theta)=\int\limits_0^{+\infty}
x^3e^{-x^2/2}L_{n-1}^{(1)}(x^2)\cos\phi         
\cos(\tau x)\,dx                   \eqno(6.1) 
        $$
$$
S_n(\tau,\theta)=\int\limits_0^{+\infty}
x^3e^{-x^2/2}L_{n-1}^{(1)}(x^2)\sin\phi           
\cos(\tau x)\,dx                     \eqno(6.2)     
        $$
$$
H_n(\tau,\theta)=\int\limits_0^{+\infty}
x^2e^{-x^2/2}L_{n-1}^{(1)}(x^2)\cos\phi                 
\sin(\tau x)\,dx                   \eqno(6.3) 
        $$
$$
G_n(\tau,\theta)=\int\limits_0^{+\infty}
x^2e^{-x^2/2}L_{n-1}^{(1)}(x^2)\sin\phi                 
\sin(\tau x)\,dx                     \eqno(6.4)      
        $$
where $\phi=1/2\,x^2\tan\theta$.

Proof of Theorem 1 essentially employs the asymptotic formula for 
Laguerre polynomials [11]     
$$
 x^3e^{-x^2}L_k^{(1)}(x^2)=\frac{1}{\sqrt \pi}
k^{\frac{1}{4}}e^{-\frac{x^2}{2}}l_k                 \eqno(6.5)   
        $$
$$
 l_k=x^\frac{3}{2}\cos\left( 2x\sqrt{k+1}-\frac{3}{4}\,\pi \right)+r_k
        $$
$$
    r_k=O\left(\frac{x^\frac{1}{2}}{\sqrt k}\right)+
    O\left(\frac{x^\frac{9}{2}}{\sqrt k}\right)
    +O\left(\frac{x^7}{k^\frac{3}{4}}\right)+h_k,\,\,\,\,\,x\ge0,\,\,\,\,\,\,  
h_k=O\left(\frac{x^\frac{3}{2}}{k^{\frac{3}{4}}}\right)
        $$
where the notation $\,O(u)\,$ means 'of the order of $\,\,u\,$'.

The following Lemma is also used [5]:

{\it Lemma.} For $-\pi < \phi <\pi$ the following formula holds
$$
s\sum_{k=1}^{+\infty}(-1)^k\frac{1}{k}L_{k-1}^{(1)}(s)
e^{ik\phi}=1-\exp\left(\frac{1}{2}s\right)\cdot
\exp\left(i\frac{1}{2}s\tan(\frac{1}{2}\phi)\right)
                                                          \eqno(6.6)    
$$
The series on the left side of (6.6) converges uniformly in any rectangle \linebreak
$\,|\phi|\le \pi -\delta,\,\,$ $0\le s\le s_0\,$ ($s_0>0$). 

{\it  Proof of Theorem 1.}
When in the series (5.3) and (5.4) the products of trigonometric
functions are decomposed into sums, the rearranged series take the form
$$
\hat C_m(\tau,\theta)=-\frac{1}{2}\sum_{k=0}^{+\infty}
(-1)^k\hat l_{km}(\tau)\cos(2k\theta)                    \eqno(6.7)     
        $$
$$
\hat S_m(\tau,\theta)=-\frac{1}{2}\sum_{k=0}^{+\infty}
(-1)^k\hat l_{km}(\tau)\sin(2k\theta)                    \eqno(6.8) 
        $$
$$
\hat l_{km}=-l_{k+1,m}+2l_{km}-l_{k-1,m},\,\,\,\,\,\,l_{-1,m}=l_{0,m}=0
        $$
or, by (5.7) and (5.8),
$$
\hat l_{0,m}=-l_{1,m}(\tau)=-2\int\limits_0^{+\infty}
x^3B(x)\,dx     
        $$                                             
$$
\hbox{for}\,\,\,k\ge 1\,\,\,\hat l_{km}(\tau)=\frac{2}{k}\int\limits_0^{+\infty}
x^5L_{k-1}^{(1)}B(x)\,dx         \eqno(6.9)      
        $$
$$
B(x)=\exp{(-x^2)}L_{m-1}^{(1)}(x^2)\cos(\tau x)
	$$
It follows from (5.7) and (6.5) that $l_{nm}=O(1/n^{1/4})$ 
 $\hbox{for}\,\,{0\le{\tau}}\le{T},\,\,$ where $\,\,T\,\,$ and 
$\,\,m\,\,$ are fixed. 

The difference of the $n$-th partial sums of series (5.3) and (6.7)
tends to zero as $n$ tends to $\,\,+\infty,\,\,$ since $\,\,l_{nm}\to 0\,\,$
as $\,\,n\to{+\infty}.\,\,$ This means, that either both series converge
to the same sum or both series diverge.  
 
It follows from (6.5) and (6.9) that $\hat l_{nm}=O(1/n^{5/4})$. 
Now we conclude that 
series (6.7) converges absolutely and uniformly throughout the region
$\,\,{0\le{\tau}}\le{T},\,\,|\theta|\le \pi/2\, $ 
and series (6.5) converges uniformly in the same region.

The same result for series (5.4) - (5.6) can be proved in a similar
manner.  This completes the proof of Theorem 1.
\bigskip

{\it Proof of Theorem 2.}
By the Lemma
$$
x^2\sum_{k=1}^{+\infty}(-1)^k\frac{1}{k}L_{k-1}^{(1)}(x^2)\cos(2k\theta)=
1-\exp\left(\frac{1}{2}\,x^2\right)\cdot
\cos\left(\frac{1}{2}\,x^2\cdot \tan \theta \right)   \eqno(6.10)    
	$$
At any fixed value of $\theta $, the series on the left side of (6.10) 
converges uniformly within any segment $0\le x\le a$. 

In the integral 
$$
\int\limits_{0}^{a}x^3e^{-x^2/2}L_{m-1}^{(1)}(x^2)
\cos\left(\frac{1}{2}\,x^2\cdot \tan \theta \right)\cos(\tau x)\,dx 
	$$
we replace the integrand with the series on the left side of (6.10) 
and perform the integration term by term. 
The integrals converge absolutely as $\,a \to+\infty\,$. 

Letting $a$ tend to $+\infty$, we obtain (6.7) and, consequently, (6.1).  

Equalities (6.2) - (6.4) can be proved in the similar way.

In the proof of Theorm 1, it was found that the right-hand parts in (6.1) - (6.4) 
drop as $n^{-1/4}$. Series (5.2) converges absolutely, if values of $a_n,\,\,$ $b_n,\,\,$ 
$\rho_n,\,\,$ and $e_n$ drops, for example, as $1/n$.

Series (5.2) converges absolutely and uniformly with regard to $ \theta $ and 
$ \tau $, if the initial free surface shape $W(\theta,0)$ and initial doublet
distribution $\nu(\theta,0)$ are smooth functions and, consequently, can be 
expanded in a convergent Fourier series. 
\bigskip

{\bf 7. Specific wave packets}
\bigskip

{\bf Theorem 3.} The following expansions hold:  
$$
C_n(\tau,\theta)=\sum_{k=1}^{+\infty}
2^k\,k!\,b_{n,k}(\tau)(\cos\theta)^{k+1}\cos((k+1)\theta)     
	$$
                           $$                \eqno(7.1) $$   
$$
S_n(\tau,\theta)=\sum_{k=1}^{+\infty}2^k\,k!\,
b_{n,k}(\tau)(\cos\theta)^{k+1}\sin((k+1)\theta)     
	$$
The coefficients $b_{n,k}(\tau)$ are defined by the following power series:  
$$
F_{n,1}\equiv x^2L_{n-1}^{(1)}(x^2) \cos(\tau x) = 
\sum_{k=1}^{+\infty}b_{n,k}(\tau)(x^2)^k     \eqno(7.2) 
        $$
In any rectangle $|\theta|\le\pi/2$, $ 0\le \tau \le T,\,\,$ 
($T>0$ may  be assigned arbitrarily)
series (7.1) converges uniformly with regard to $ \theta $ and $ \tau $. 

Similar expansions hold for $H_n(\tau,\theta)$ and $G_n(\tau,\theta)$. 
\bigskip

{\it Proof of Theorem 3.} Write 
$$
C_m=\int\limits_0^{+\infty}F_{m,1}\pd {V_1}{x}\,dx,\,\,\,\,\,\,
\pd {V_1}{x}= xe^{-x^2/2} \cos\phi    
	$$
The function $F_{m,1}$ is defined by (7.2). In any rectangle 
$0\le x\le a$, $0\le \tau\le T$ series (7.2) converges uniformly 
with regard to $x$ and $\tau$.

Integration by parts gives
$$
C_m=\cos\theta\cdot \int\limits_0^{+\infty}F_{m,2}\pd {V_2}{x}\,dx
	$$
where 
$$
xF_{m,2}=\pd {F_{m,1}}{x},\,\,\,\,\,\, 
F_{m,2}=2\sum_{k=1}^{+\infty}kb_{m,k}(\tau)(x^2)^{k-1}
	$$
$$
\pd {V_2}{x}= xe^{-x^2/2}\left[\sin\theta \cos\phi+   
\cos\theta \sin\phi\right]     
	$$
$$
V_2=-\cos\theta e^{-x^2/2}
\left[\sin(2\theta) \cos\phi+    
\cos(2\theta) \sin\phi\right] ,\,\,\,\,\,\,    
\phi=1/2\,x^2\tan\theta
	$$

Performing the integration by parts again and again we come to the first 
equality (7.1).
The second equality (7.1) can be proved in a similar manner.

It is not difficult to find that at any fixed value of $m$, coefficients 
$2^kk!b_{m,k}$ go to zero fast enough as $k\rightarrow +\infty$ 
to assure uniform convergence of series (7.1).

For $m=1$ we have 
$$
F_{1,1}=x^2\cos(\tau x)
	$$
$$
2^kk!b_{1,k}= 
(-1)^{k-1}\frac{2k}{(2k-3)!!}(\tau^2)^{k-1}\,\,\,\,\,(k=1,\,2,\,\cdots )
	$$
$$
(2k-1)!!=1\cdot 3\cdot 5\cdot\dots (2k-1),\,\,\,\,\,\,(-1)!!=0
	$$
If $m=2$, then
$$
F_{2,1}=x^2(2-x^2)\cos(\tau x)
	$$
$$
2^kk!b_{2,k}=
(-1)^{k-1}\frac{4k}{(2k-1)!!}(\tau^2)^{k-1}+
(-1)^{k-1}\frac{4k(k-1)}{(2k-3)!!}(\tau^2)^{k-2}
	$$

Expression (7.1) shows that $W_0(\theta, t)$ satisfy boundary condition 
(3.5).

If in (7.1) coefficient $b_{m,1}$ does not equal zero, $W_0(\theta,t)$ 
drops as  $1/y^2$ when $y$  increases without bound. But the departure 
of the interface from its equilibrium position may drop faster: 
for any given natural number $n$ and suitable initial conditions 
the departure drops as $1/y^{n+1}$ (and even faster). 
Indeed, there exist such value of $\lambda_k$ that
$$                          
u^n=\sum_{k=0}^n\lambda_kL_k^{(1)}(u)
	$$
so at initial conditions in (5.2) corresponding  to $\lambda_k$ we obtain 
$W_0(\tau,\theta)=I_{2n+1},$
$$ 
 I_{2n+1}(\tau,\theta)=\int\limits_0^{+\infty}
x^{2n+1}e^{-x^2/2}\cos\phi\cos(\tau x)\,dx   
	$$
In this case, expansion (7.2) should be replaced by
$$
F(x)\equiv x^{2n}\cos(\tau x) = 
\sum_{k=n}^{+\infty}b_k(\tau)(x^2)^k  
        $$
and, consequently, 
$$
 I_{2n+1}(\tau,\theta)=\sum_{j=0}^{+\infty}(-1)^j\,2^{n+j}\cdot (n+j)!\,
\frac{\tau^{2j}}{(2j)!}
(\cos\theta)^{n+1+j}\cos((n+1+j)\theta)
	$$
$$
 I_{2n+1}(0,\theta)=2^n\cdot n!\,(\cos\theta)^{n+1}\cos((n+1)\theta)
	$$
The integral is normalized by   $2^n\cdot n!$:
$$
I_{2n+1}(\tau,\theta)=2^n\cdot n!\,\hat I_{2n+1}(\tau,\theta),\,\,\,\,\,\,
\hat I_{2n+1}(0,0)=1 
	$$
$$
\hat I_{2n+1}(0,\theta)=(\cos\theta)^{n+1}\cos((n+1)\theta)
	$$
Functions $C_m(\tau,\theta)$ are linear combinations of integrals 
$I_{2n+1}(\tau,\theta)$; for instance, $C_1(\tau,\theta)=I_3(\tau,\theta),\,$ 
$C_2=2I_3-I_5,\,$ $2C_3=2I_7-6I_5+6I_3$. 

Similar results for $S_m(\tau,\theta)\,$, $H_m(\tau,\theta)\,$, and  $G_m(\tau,\theta)$ 
can be obtained in the similar way.

We define spescific wave packets by equations 
$$ 
x=c_nI_{2n+1},\,\,\,\, y=(x-f)\tan\theta 
	$$
$$
I_{2n+1}(\tau,\theta)=\int\limits_0^{+\infty}
x^{2n+1}e^{-x^2/2}\cos(\frac{1}{2}x^2\tan\theta)\cos(\tau x)\,dx;    \eqno(7.3)   
	$$
$n$ is referred to as the packet number.

Equations of other three sets of wave packets will be obtained from (7.3) if 
the product of trigonometric functions in the integrand is replaced 
by one of the following expressions
$$
\cos\phi\sin(\tau x)/x,\,\,\,\,\,\,
\sin\phi\cos(\tau x),
	$$
$$
\sin\phi\sin(\tau x)/x,\,\,\,\,\,\,\phi=\frac{1}{2}x^2\tan\theta
	$$

By equations (6.1), (6.2) and three Theorems, any wave group on the free 
surface is a mixture of finite or infinite (it depends on initial conditions) 
set of the specific wave packets of different numbers,  and evolution of each packet 
in the mixture is not influenced by evolution of the others. 

Zeros of the waves are defined by the equation $I_{2n+1}(\tau;\theta)=0$ and 
(at fixed value of $\tau$) are situated in the numbered rays 
$\theta=\theta_k(\tau)$ ($k$ is the number of a zero). 
For brevity we will use the term 'zero $\theta(\tau)$' to denote a zero 
of a wave.

Zeros $\theta_k(\tau)$ of the wave packet (7.3) are independent of $f$ 
and $c_n$.

Horizontal coordinate of the zero $\theta_k(\tau)$ is given 
by $y_k=-f\cdot \tan\theta_k(\tau)$. This leads to the following 

{\bf Assertion 1}: At any given value of $\tau$

i) the vertical coordinates of the crests and the troughs (and, consequently, 
the waveheights) are independent of $f$;

ii) the ratio of the distances $y_{k+1}(\tau)-y_k(\tau)$ and
$y_k(\tau)-y_{k-1}(\tau)$ between any successive zeros of the waves is 
independent of $f$.

In the course of time the number of zeros of the waves  
increases so the existence of the functions $\theta_k(\tau)$ on the whole
semiaxis $\tau>0$ is not assured. 

Horizontal velocity of a zero of the wave is $d y_k/d t=u_k(\tau)\sqrt{0.5|f|}$, 
where $u_k(\tau)=d\tan\theta_k(\tau)/d\tau$.

The definition does not guarantee the existence of
differentiable (and even continuous) functions $\tan\theta_k(\tau)$, 
but the average rate
$$
u_k=[\tan\theta(\tau+\Delta\tau)-\tan\theta(\tau)]/\Delta\tau
	$$
exists, and this gives us a useful piece of information about the speed 
of waves as we will see in section 10.

Let $L_*$ be the dimensional unit of length, then $T_*=\sqrt{L_*/g}\,$ is 
the dimensional unit of time.

At the instant $t_*$ the dimensional coordinate 
of the zero $\theta=\theta(\tau)$ is 
$$
y_*(t_*)=|f|L_*\tan\theta(\tau), \,\,\,\,\,\,
t_*=\tau\sqrt{2|f|}\cdot T_*
	$$ 	
and, consequently, the ratio 
$$
\lambda(\tau)=\frac{y_*(t_*)}{gt_*^2}=
\frac{1}{2}\,\,\frac{\tan\theta(\tau)}{\tau^2}      \eqno(7.4)   
	$$
depends on $\tau$ only. 

 Given a fixed value of $\tau$, $\tan\theta(\tau)$  can be calculated from equations of a wave packet, and corresponding value of $\lambda(\tau)$ can be obtained from (7.4). 
Calculations show that $\tan\theta$ and $\lambda$ are monotone functions of $\tau$, so 
$\tau$ and $\tan\theta$ can be calculated for given value of $\lambda$.

At any particular moment of time, the system of waves contains only one 
wave of maximum height (WMH) on semiaxis $y>0$  
(the situation with two waves of equal maximum height can be ignored), so
the zeros of WMH constitute a 'natural frame of reference' for other zeros. 

The wave of maximum height is singled out by three zeros and consists 
of a crest and the trough following or preceding the crest.
Let $\theta_r(\tau)$ and $\theta_f(\tau)$  denote two of the three zeros 
of the WMH which correspond to minimum and maximum of the three 
$|\tan\theta(\tau)|$ respectively. By the zero $\theta_f(\tau)$ we define 
the front of WMH at the instant $\tau$, by $\theta_r(\tau)$ the rear 
of the wave is determined.  
\bigskip

{\bf 8. Evolution of specific wave packets} 
\bigskip

The figures 2 - 6 illustrate evolution of wave profiles specified by equations 
(at $f=-10$)
$$
x=0.4\,\hat I_{23}(\tau,\theta)+0.075\,\hat I_5(\tau,\theta),\,\,\,\,\,  
|y|=(x-f)|\tan\theta|                                              \eqno(8.1)  
	$$
$$
x=0.4\,\hat I_{23}(\tau,\theta),\,\,\,\,\,|y|=(x-f)|\tan\theta|    \eqno(8.2)   
	$$
$$
x=0.075\,\hat I_5(\tau,\theta),\,\,\,\,\,|y|=(x-f)|\tan\theta|  \eqno(8.3)   
	$$

\begin{figure}
	\resizebox{\textwidth}{!}
		{\includegraphics{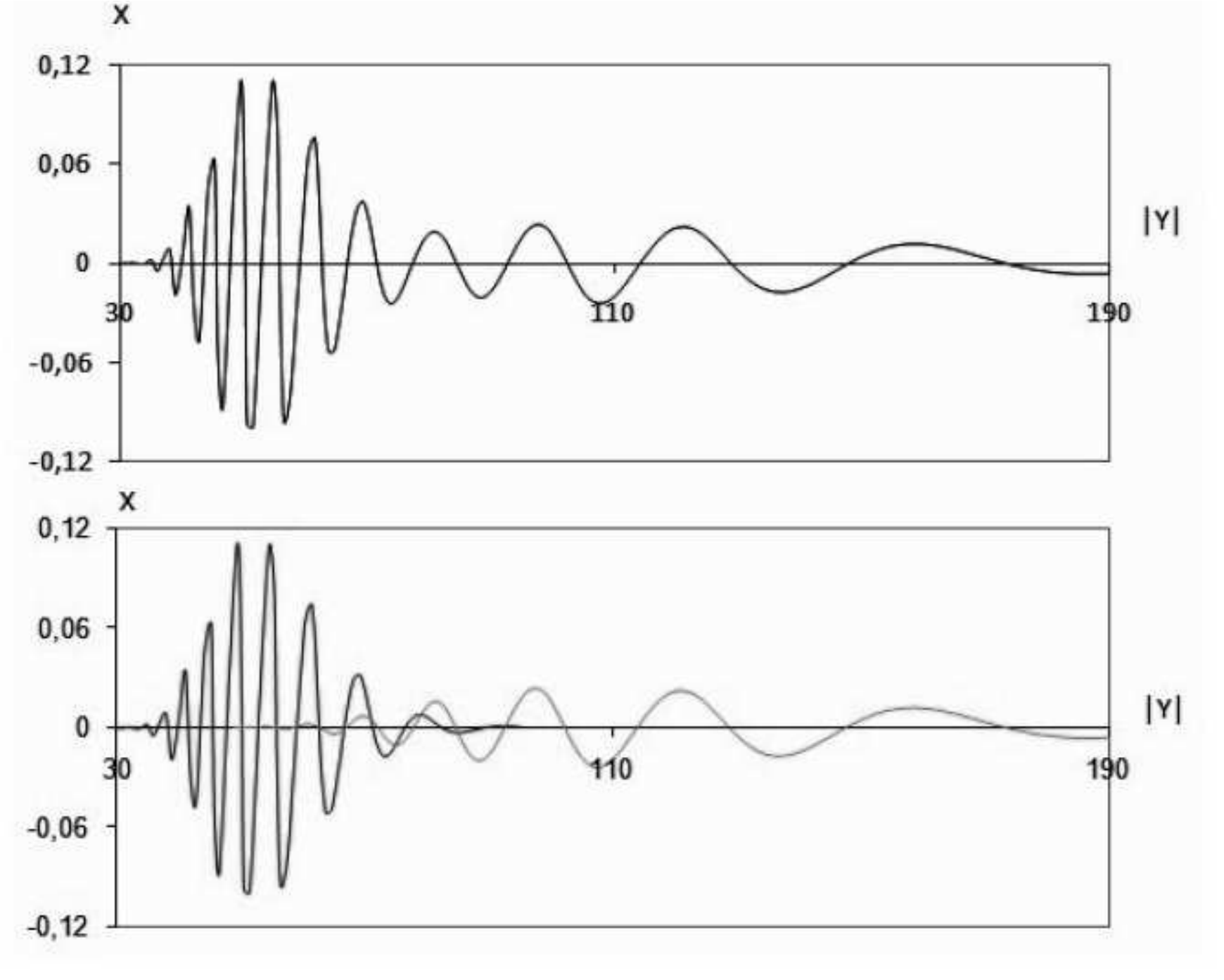}}  
	\caption{
	Instantaneous profiles of the waves (8.1) (upper panel), 
(8.2) (bottom panel, the curve on the left), 
and (8.3) (bottom panel, the curve on the right) at $\tau=25$.  
	}
\end{figure} 

\begin{figure}
	\resizebox{\textwidth}{!}
		{\includegraphics{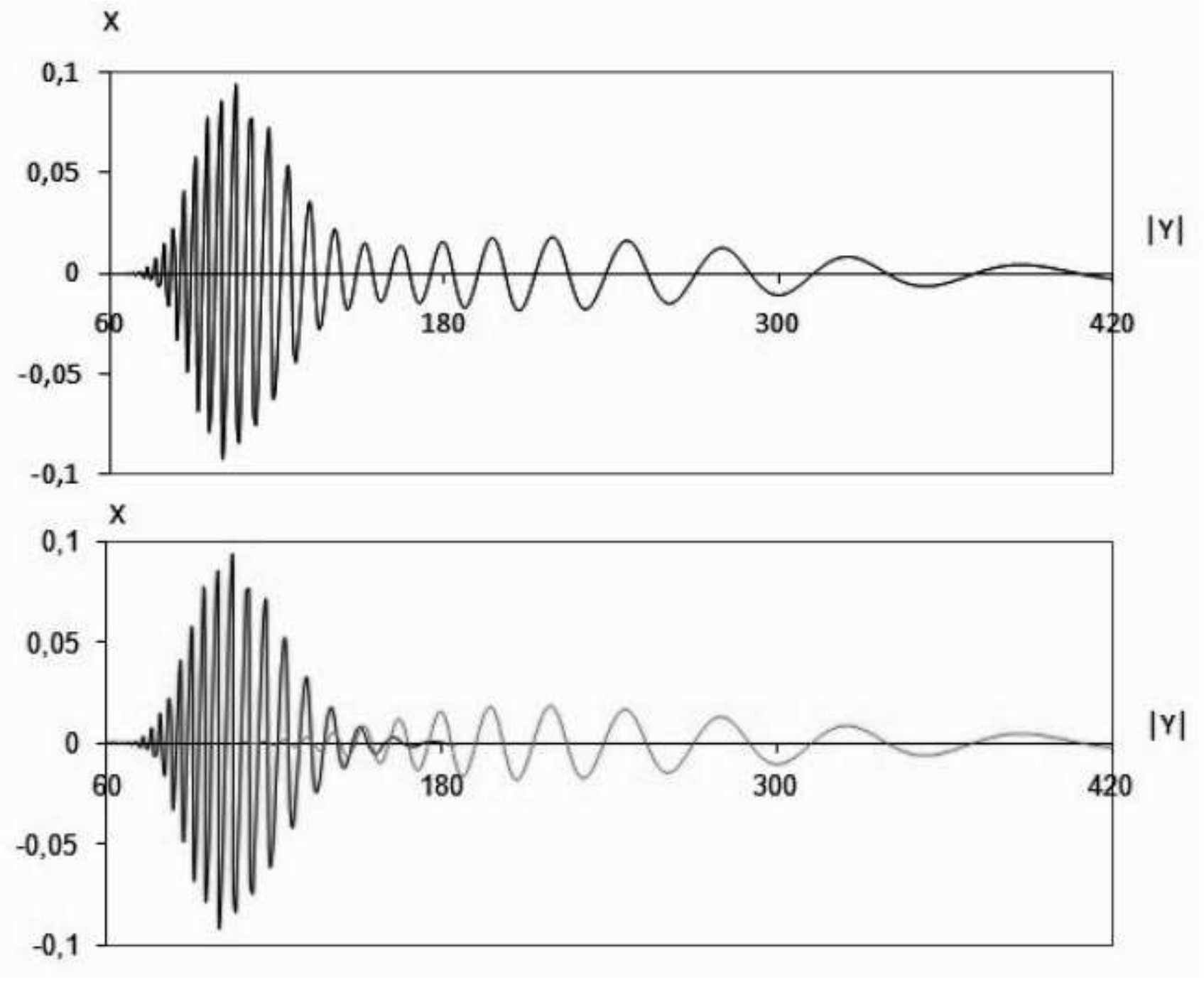}} 
	\caption{
  Same as Figure 2, but $\tau=50$.
	}
\end{figure} 

\begin{figure}
	\resizebox{\textwidth}{!}
		{\includegraphics{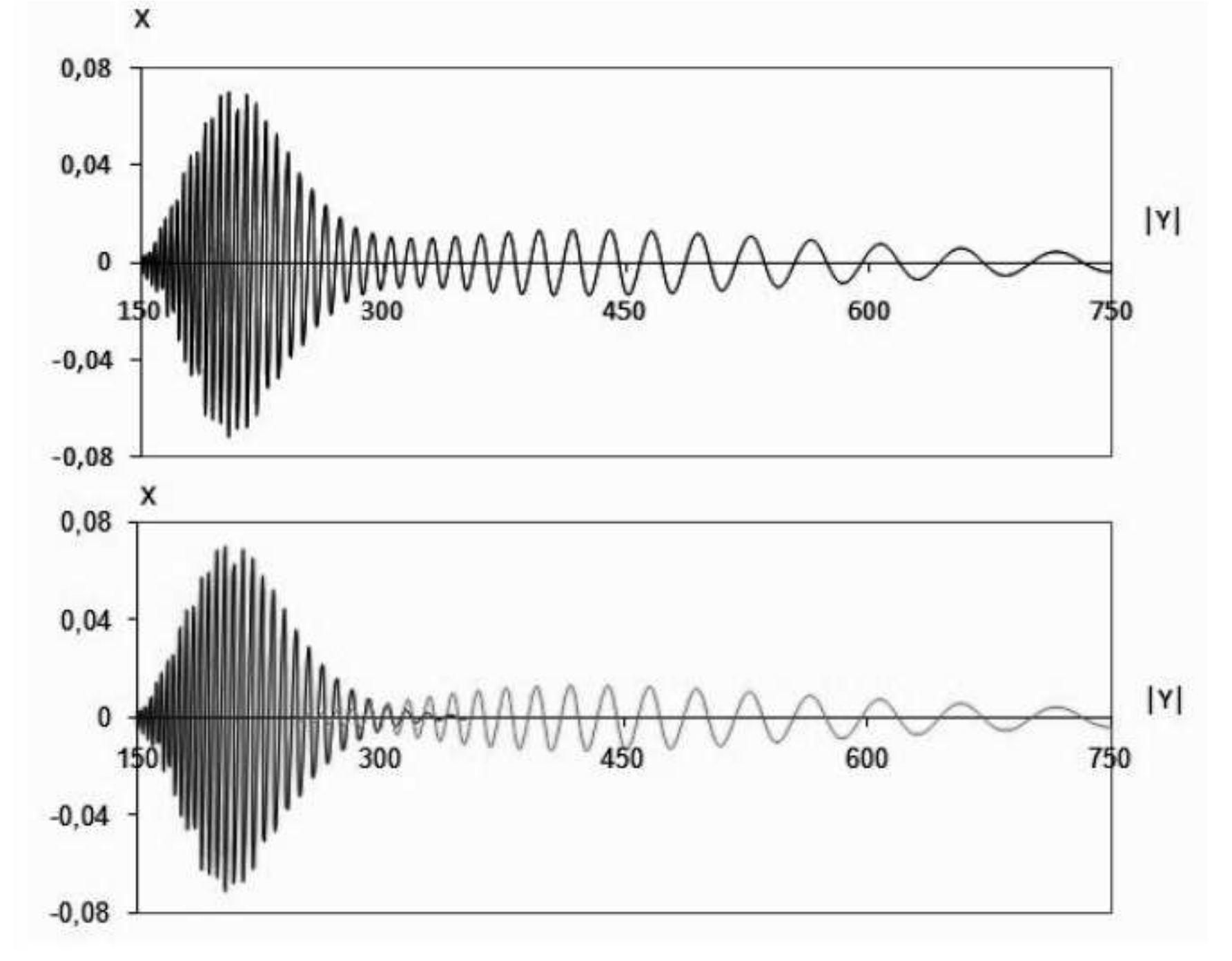}} 
	\caption{
  Same as Figure 2, but $\tau=100$.
	}
\end{figure} 

\begin{figure}
	\resizebox{\textwidth}{!}
		{\includegraphics{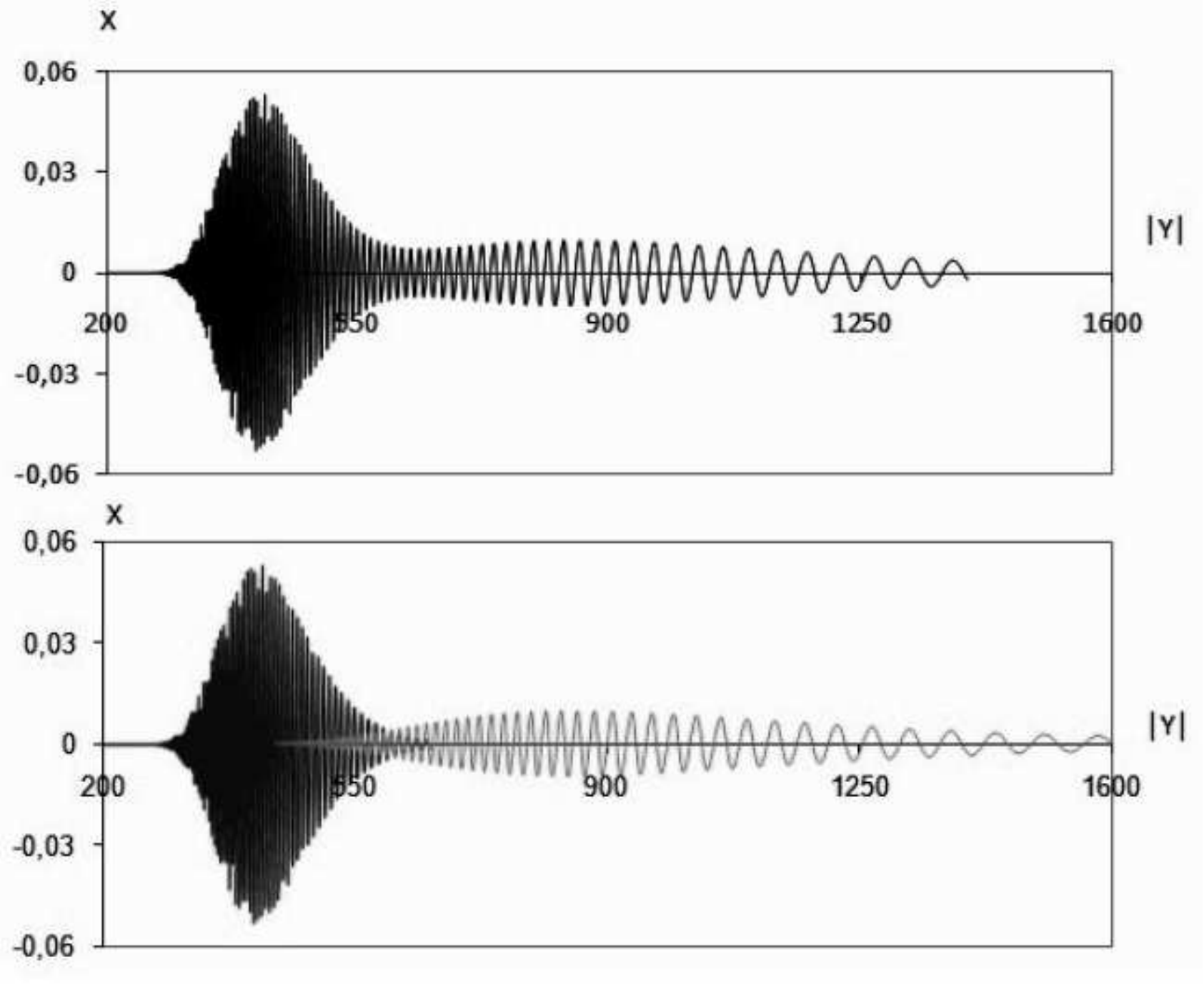}}
	\caption{
  Same as Figure 2, but $\tau=200$.
	}
\end{figure} 

\begin{figure}
	\resizebox{\textwidth}{!}
		{\includegraphics{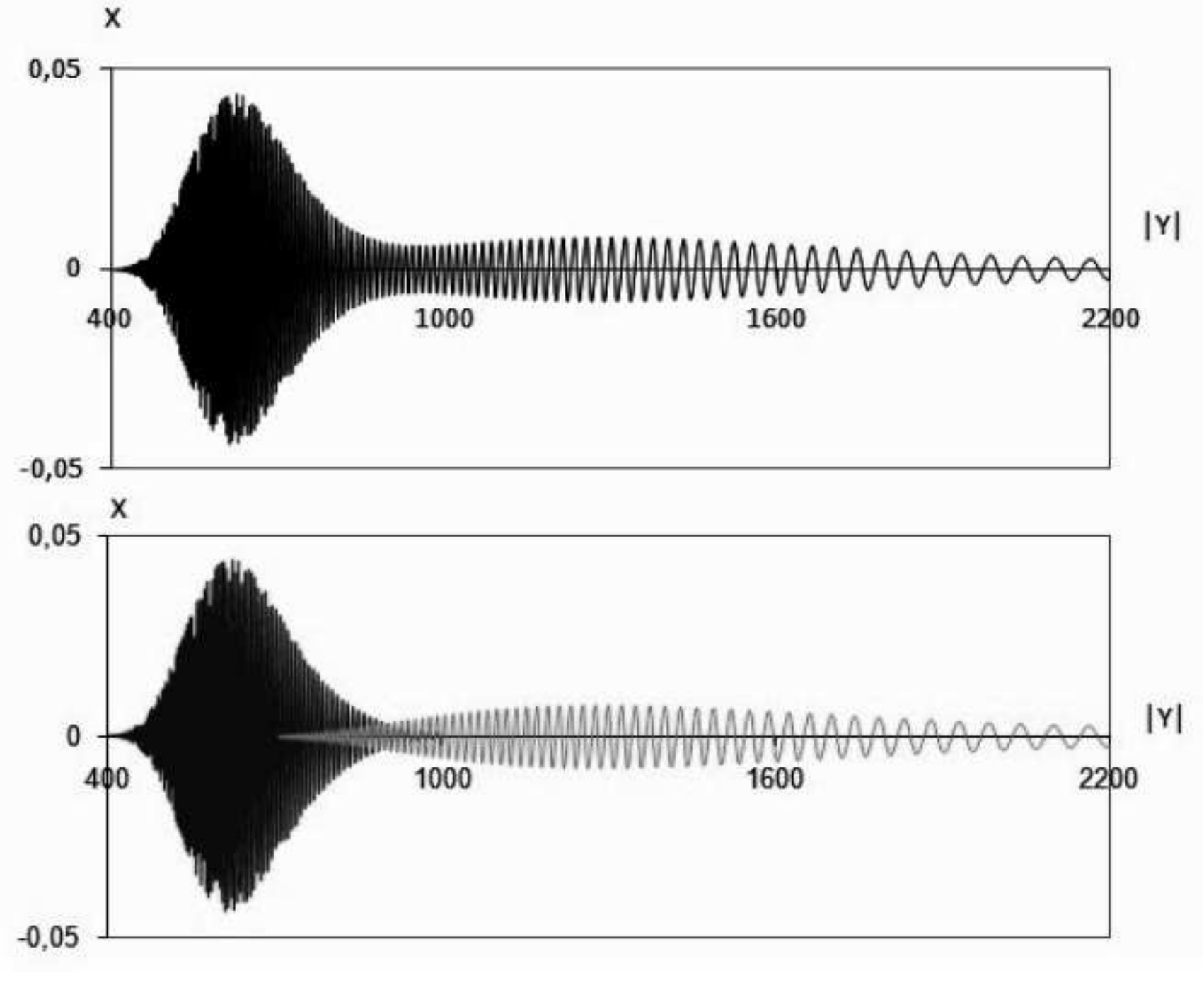}}
	\caption{
  Same as Figure 2, but $\tau=300$.
	}
\end{figure} 
At each value of $\tau$, the profiles are drawn in the same scale. 
For each of the profiles, initially,jpgjpg at $\tau=0$, the crest of the wave 
of maximum height is situated on the ray $\theta=0$, i.e., on the $x$-axis; 
the vertical coordinates of the crests are: $x=0.4$ for (8.2), 
$x=0.075$ for (8.3), and, consequently, $x=0.475$  for (8.1).

The figures show that the initial free surface displacement turns into two 
wave groups which run away from $x$-axis in opposite directions. From the 
figure we notice that with time the waves leave a region about 
the vertical plane $y=0$. As there is no flow across the plane, 
the fluid motion would be unaffected if the plane was replaced by 
rigid barrier. From now on we will consider wave groups propagating 
in a fixed direction, say, along the semiaxis $y>0$.

We note in passing that, in case of plane waves, the term 'wave group' 
means a system of waves in which the amplitude dies away on either side of 
the wave of maximum height; the term 'wave'\, means a
wave-group's section which consists of one crest and the adjacent trough and 
is singled out by three successive zeros of the group;
the level difference between a crest and the following or the preceding trough  
is referred to as the waveheight (or height of the wave); 
the term 'wave packet' means a wave group in which 
the amplitude dies away very quickly with distance from the wave of maximum 
height; figures 2 and 3 shows, that the waveheight of the wave of maximum height 
of the wave group (8.2) is about 0.21 at $\tau=25$ and 0.18 at $\tau=50$.

System of harmonic waves $x=\cos(k y-\omega t)$ on the free surface infinite 
in extent is not a wave group.

The wave group (8.1) is a nonlinear combination of wave packets (8.2) and (8.3). 
The figures show that the wave packets (8.2) and (8.3) propagate at different 
speeds: the horizontal distance between the crests of the waves of maximum 
height is approximately 70 at $\tau=25,\,$ 115 at $\tau=50,\,$ 260 at $\tau=100,\,$ 
440 at $\tau=200,\,$ and 690 at $\tau=300$.

The wave group (8.1) gradually disintegrates into its components (8.2) and 
(8.3), and each of the components doesn't affect the evolution of the another 
one. 

The greater the number of a packet, 
the shorter the wavelength of the packet's carrier wave component and  
the slower the packet travels. 
The wavelength and velocity of simple harmonic waves on deep water 
are related in a similar way.
\bigskip

{\bf 9. Infinitesimal linear wave packets: limit case of the nonlinear theory}
\bigskip

Setting $y_1=y/f$, substituting $y_1f/(x-f)$ 
for $\tan\theta$ in $\hat I_{2m+1}$, 
letting $|f|$ tend to $+\infty$, we obtain 
the equations of the packet with number $m$ as
$$
x=W(y,t),\,\,\,\,\,\,
W=c\int\limits_0^{+\infty}
s^{2m+1}e^{-s^2/2}\cos(\frac{1}{2}s^2y)\cos(t s)\,ds   
	$$
where $\tau$ has been replaced by $t$ and subscript 1 has been dropped.

Calculations show that the leading-order term $\Phi_0(x,y,t)$ for velocity
potential and the function $W(y,t)$ satisfy equations
$$
\pd {W(y,t)}{t}=\pd {\Phi_0}{n_-},\,\,\,\,\,\,\,\,\,\,\, \pd {\Phi_0}{t_-}+W=0   
                                       \eqno(9.1)       
 $$
where subscript "n" is used for the derivative in normal direction to the free surface  $x=W(y,t)$, and equations (9.1) involve
 the partial derivatives at point $x,y$ on the free surface;
$t$ is the non-dimensional time introduced in subsection 1.2.

By the way, note that at any value of $f$ equation (3.3) may be rewritten as 
$$
\pd {W}{t}=\pd {\Phi}{n_-}
	$$

Equations (9.1) are the boundary conditions (ii) and (iii) named in the Introduction. 
If the conditions are moved from the free surface to the equilibrium plane
$\,x=0,\,$ the normal derivative reads $\partial \Phi_0/\partial x|_{x=0}$.

We come to the following 

{\bf Assertion 2}:

As a limit case of the nonlinear theory,  equations for infinitesimal deep water waves are obtained with boundary conditions on the evolving free surface, and the solution to the equations is found.  
If the boundary conditions are shifted from the evolving free surface to the equilibrium position of the surface, the equations reduce to equations of classical linear theory of infinitesimal waves. 
\bigskip

{\bf 10. Testing of the theory against experiments performed in a water tank} 
\bigskip

We use some results from wave pulse experiments presented by 
Feir [9] and Yuen \& Lake [8] bearing in mind that 
the aim of Feir is to study the effects of finite amplitude of a wave group on 
the group shape and the frequency distribution over the group
while the main purpose of the work by Yuen \& Lake is to 
decide whether the predictions based on the Schr\"odinger equation correspond 
with experimental results.

The wave groups were generated in a tank equipped with a wavemaker. 
In [9] each wave pulse was recorded at one fixed point of the tank, 
while in [8] wave gauges located at the distances 5 ft, 
10 ft, 15 ft, 20 ft, 25 ft, and 30 ft downstream of the wavemaker were used 
to record a wave group on the water surface so that the evolution of the wave group 
can be demonstrated. 

The figure 7 is a reproduction of figure 1 shown in [8]. 
 
\begin{figure}
	\resizebox{\textwidth}{!}
		{\includegraphics{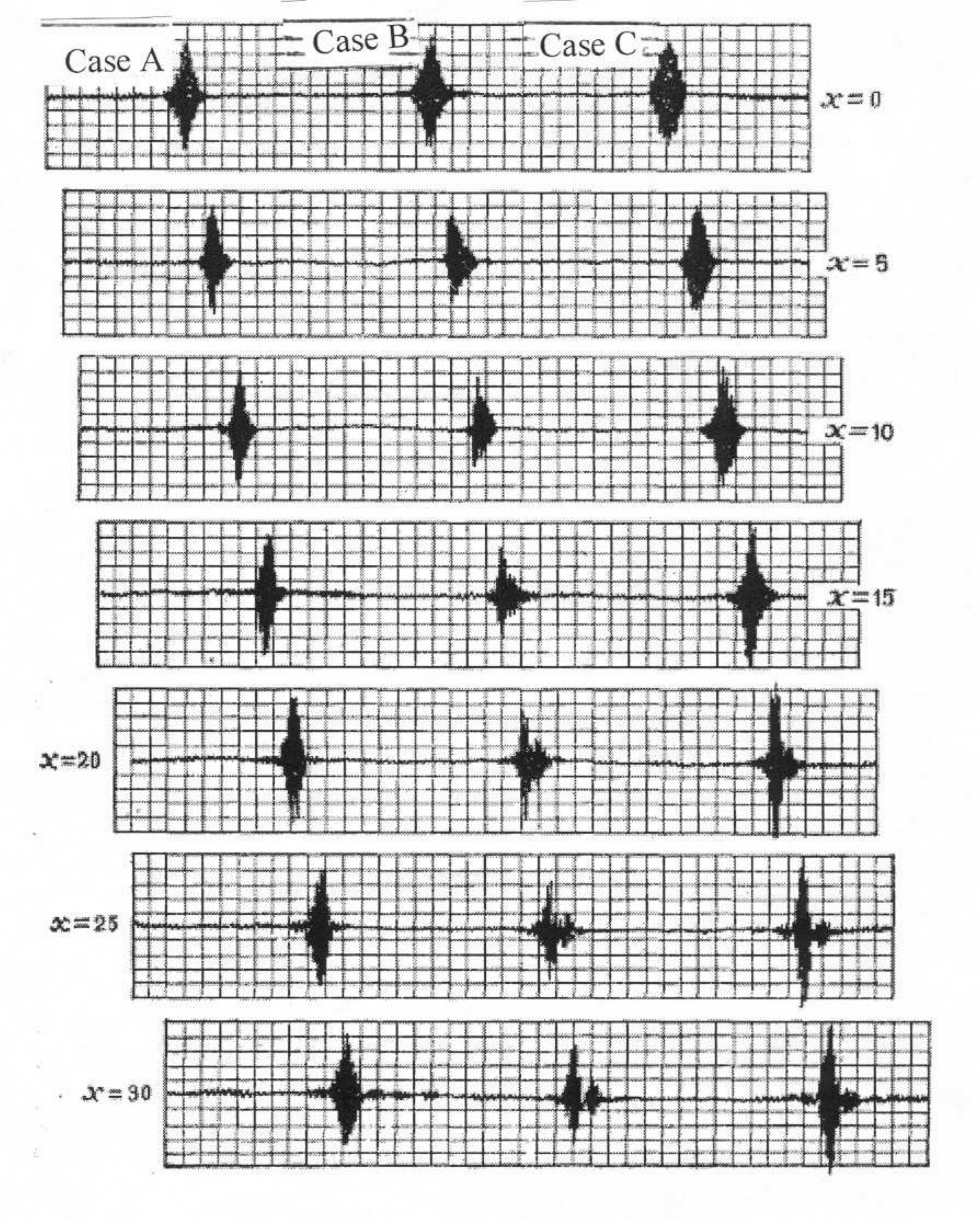}} 
	\caption{
  Evolution of three wave envelope pulses (reproduced from [8]).
	}
\end{figure} 

Evolution of three wave envelopes is shown in cases $A,\,$ $B\,\,$ and $C$ 
cor\-res\-pon\-ding to three initial profiles (given at $x=0$) symmetric about 
their peaks: ($A$) envelope soliton, ($B$) a  hyperbolic secant envelope, 
($C$) a sine envelope of the same amplitude as in the case $A$; the amplitude 
scale of the hyperbolic secant envelope is reduced by a factor of 2.5 compared 
to the case $A$; in the three cases the carrier frequency of the wavemaker was 
$2\,$ Hz; the experiments were performed in a water tank 3 feet deep. 
 In figure 7 the wave envelopes propagate along the tank from the 
left to the right.

During the time interval of the observation, in case $A$, the shape of the 
envelope in the figure seems to be almost unchanged with time; in cases 
$B$ and $C$, initial wave group gradually turns into a structure which consists 
of a relatively 'tall' wave packet following a 'low' one and long chains 
of waves of small amplitude in the front and in the rear of the system of 
packets.
Formation of similar structures is shown in [8].   

In [8] experimental measurements are not presented, 
actual dimensions of the envelopes as well as the ratio of vertical 
to horizontal scales and the individual wave profiles are not shown.

We can obtain an idea about the horizontal and vertical scales only 
for the initial envelope soliton (figure 7, case $A$, $x=0$).

The shape of the envelope soliton is given by 
$x=a\,{\rm sech}\,u,\,$ $u=\sqrt{2}k^2a\,s$, $ka\approx 0.14$; 
$s=0$ is the vertical axis of symmetry of the shape, 
$a$ is the height of the envelope;  maximum slope of the tangent 
to the envelope is reached at $u=0.882$; coordinates of the point of tangency 
are $s=32.14a,\,\,\,x=0.7a$; at the level of the point, i.e., on the level 
of 0.7 of the envelop's height, the width of the envelope is nearly 64 times 
as great as the height.

But in figure 7, case A, at $x=0$, the envelope's width on the level 
0.7 of it's height is approximately of 0.2 of the height, so the ratio of 
the vertical scale to the horizontal scale is about 300.

Figure 8 shows the same profiles as the figures 2 - 6, but the horizontal 
scales are reduced significantly.   

\begin{figure}
	\resizebox{\textwidth}{!}
		{\includegraphics{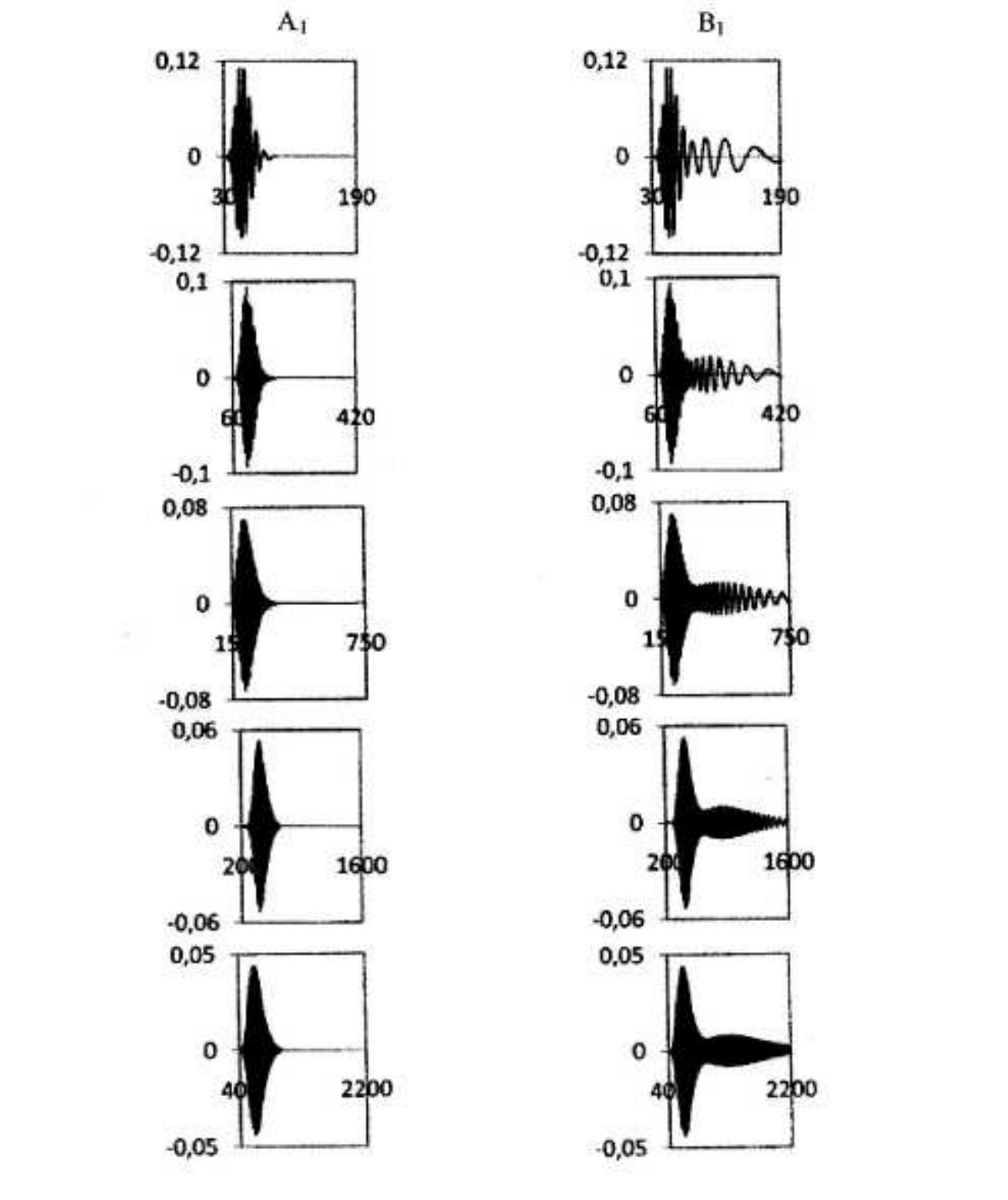}} 
	\caption{
  Same as figures 2 - 6 (top to bottom) at $\tau=25,\,$ 
$\tau=50,\,$ $\tau=100,\,$ $\tau=200,\,$ $\tau=300\,$ respectively, 
but on reduced horizontal scale.  Column $A_1$: profiles of packet (8.2),  
column $B_1$: profiles of packet (8.1).
	}
\end{figure} 

In figure 8, column $A_1$, 
the envelopes seem to be of the same height and the same width.
The resemblance between columns $B$ (figure 7) and $B_1$ (figure 8) is 
doubtless.

At $\tau=25$ (figure 2, bottom panel, the curve on the left) the height of 
the packet (8.2) equals 0.1136, the packet's width at the level 0.7 of its 
height is equal to 6.6, 60 times as great as the height (as for envelope 
soliton with $ka=0.144$). The length of the packet increases with time, while 
the height decreases (figures 2 - 6, bottom panels, the curves on the left; 
formation of a chain of waves of small amplitude in the front of the packet is 
demonstrated in the top panels). 

In [9], a number of records of wave pulses are shown generated by 
a wave\-ma\-ker oscillating at a carrier frequency of 2.5 Hz with amplitude 
varying slowly from zero to some maximum and back to zero; the range of 
the maximum amplitude of the wavemaker stroke was from 0.19 cm for the wave 
pulses of small amplitude to 1.5 cm for the pulses of large amplitude; 
in the records the range of maximum water surface displacement is from 1 mm 
to 10 mm. 
The maximum displacement of 6 mm at the distance of 8.53 m from the wavemaker 
is considered to be large (figure 2, p. 55.)

 In [9], figure 1, p. 54, the record of a wave pulse was taken 
at the distance  8.53 m from the wavemaker. The upper part of the figure 1 
shows the group of 25 individual waves which passes the gauge for 10 s.
This means  that the mean period of the waves is 0.4 s which equals to the 
wavemaker period. In the record, maximum water surface displacement is about 
1 mm. Periods of the individual waves were measured and the results are shown 
in the lower part of the figure 1: in the figure, periods of 18 waves in 
the middle of the wave pulse are close to 0.4 s; periods of the waves at 
the ends of the pulse differ from the wavemaker period up to 20\%.

Profile of packet (8.2) at $\tau=100$ (figure 4, bottom panel, the curve 
on the left) is similar to the wave profile shown in [9], figure 1, p. 54: in figure 4, bottom panel, $160<y<300$, we can clearly see 25 individual 
waves. 

The front of the wave of maximum height of packet (8.2) travels a distance of 
$y_*$ from the origin $y=0$ in a time $t_*$. Below for specific values of 
$y_*$, $\,t_*$ and  initial height of the packet, $H_*$, the following 
dimensional parameters of the packet are estimated:

$L_*\,\,$ - the dimensional unit of length (see section 3);

$h_*(t_*)$ - the maximum height of the packet at the instant $t_*$; 

$V_*\,\,$ - the average speed of the packet during the travel time $t_*$; 

$v_*\,\,$ - the speed of the WMH at the instant $t_*$; 

$l_*\,\,$ - the length of WMH at the instant $t_*$; 

$l_{**}\,\,$ - the length of the packet at the instant $t_*$.

Algorithm for calculating values of the parameters may be summarized as follows: 

Calculate the value of $y_*/t_*^2$ and then find the front of the wave 
of maximum height of packet (8.2) such that $\tau$ and $\tan\theta_f(\tau)$ 
satisfy equation (7.4); also find $\tan\theta_r(\tau)$, non-dimensional 
height $h(\tau)$ of the WMH, 
and $\tan\theta_f(\tau+\Delta\tau)$ at small $\Delta\tau$. 

From equations $H_*=c\,L_*$ and $|f|L_*|\tan\theta_f(\tau)|=y_*\,\,$ find 
dimensional estimates for vertical $L_*$ and horizontal $a=|f|L_*$ parameters 
of length ($c$ is the non-dimensional height of the packet at $t=0$). 
The height and the length of the WMH at the instant $t_*$ are estimated as 
$h_*(t_*)=h(\tau)L_*$ and \linebreak
$\,\,l_*=a[|\tan\theta_f(\tau)|-|\tan\theta_r(\tau)|]$ respectively. 
The speeds are estimated by  $V_*=y_*/t_*$, 
$\,v_*=\sqrt{0.5ag}\,[|\tan\theta_f(\tau+\Delta\tau)|-|\tan\theta_f\tau)|]/ \Delta\tau$.
Technique for estimating the length of the packet is shown below in example 1. 

The estimates of packet parameters are independent of $c$ (in (8.2) $c=0.4$) 
and the estimate of $a$ is independent of $H_*$.
+
In Feir [8] the lengths of the individual waves and the length of the wave  
pulse shown in figure 1 was not estimated, the travel time $t_*$ 
from the wavemaker to the gauge and the maximum amplitude $H_*$ of the pulse 
are not shown. It is reasonable to think that $H_*$ is comparable with the 
maximum wavemaker stroke, and $t_*$ is greater than 10 s, but not very.

{\bf Examples.}
 
{\it Example 1.} At $y_*=8.53\,\,$m,  $t_*=27.4\,\,$s, $H_*=0.75\,$cm, 
$g=9,8\,\,\hbox{m/s}^2,\,\,$  $2y_*/(gt_*^2)=0.0023$ and from equations (8.2) we obtain 
$$
\tau=91,\,\,\,\,\tan\theta_f(91)=19.150,\,\,\,\,\tan\theta_r(91)=18.610,\,\,\,\,\tan\theta_f(91.1)=19.192 
	$$
$$
\,\,\,\,h(91)=0.075,\,\,\,\,L_*=1.86 \,\,\hbox{cm},\,\,\,\,a=|f|L_*=0.445\,\,\hbox{m}
	$$
$$
h_*=1.4 \,\,\hbox{mm},\,\,\,\,l_*=0.24\,\,\hbox{m},\,\,\,\,
V_*=0.311\,\,\hbox{m/s},\,\,\,\,v_*=0.620\,\,\hbox{m/s}
	$$
An estimate of instantaneous length of a wave packet based on a record of the 
packet depends on sensitivity of the wave amplitude gauge used for the record.

At the instant $t_*$ the body of the packet is bounded by 
$y_{min}=|f|L_* \tan\theta_{min}$ and $y_{max}=|f|L_* \tan\theta_{max}$ where 
$\tan\theta_{min}$ and $\tan\theta_{max}$ are to be found from equations 
(8.2) at obtained value of $\tau$. The body of a packet includes waves 
with amplitudes greater than a preassigned level which depends on the gauge 
sensitivity. 

Inequality $|x|>0.01$ corresponds to waves with amplitude greater than 0.2 mm. 
At the instant $t_*$ the waves are located in the interval $14.3\,a<y<25.9\,a$ 
(at $\tau=91\,\,$ $\tan\theta_{min}=14.3,\,$ $\,\tan\theta_{max}=25.9$). 
The length of the interval is estimated as 
$l_{**}=(25.9-14.3)\cdot 0.45\,\,\hbox{m}=5.2\,\,$m. 

In the interval $15\,a<y<25\,a$ the inequality $|x|>0.02$ holds. 
At the instant $t_*$ the length of the interval is estimated as 
$l_{**}=10\cdot a=4.5\,\,$m. 

At the instant $t_*$ in the interval $15.5\,a<y<22.3\,a$ 
inequality $|x|>0.03$ is satisfied, so amplitudes of the waves in the interval 
are greater than 0.6 mm. The length of the interval is estimated as 
$l_{**}=6.8\cdot a=3.0\,\,$m. 

Depending on the preassigned amplitude level, one of the three values may be 
taken as the estimate of the packet length at the instant $t_*$.

The above estimates of the average speed of the packet 
and speed of the wave of maximum height 
equal to estimates of group and phase velocities respectively based on 
linear theory for waves of infinitesimal amplitude: 
at phase velocity of 0.62 m/s of sinusoidal carrier wave of a packet  
the carrier frequency is \linebreak
$\omega=15.8\,\,\hbox{s}^{-1}$,  
the wavenumber is $k=\omega^2/g= 25.47\,\hbox{m}^{-1}$, 
the group velocity is $v=0.5\omega/k=0.31\,\,$ m/s.

{\it Example 2.} At $y_*=8.53\,\,$m,  $t_*=24\,\,$s, $H_*=0.5\,$cm, 
$g=9.8\,\,\hbox{m/s}^2$, \linebreak
 $2y_*/(gt_*^2)=0.0030$ we obtain 
$$
\tau=70,\,\,\,\,\tan\theta_f(70)=14.765,\,\,\,\,\tan\theta_r(70)=14.227,
\,\,\,\,\tan\theta_f(70.1)=14.807
	$$
$$
h(70)=0.084,\,\,\,\,L_*=1.25 \,\,\hbox{cm},\,\,\,\,a=|f|L_*=0.578\,\,\hbox{m}
	$$
$$
h_*=1.05 \,\,\hbox{mm},\,\,\,\,\,\,\,\,l_*=0.31\,\,\hbox{m},\,\,\,\,
V_*=0.355\,\,\hbox{m/s},\,\,\,\,v_*=0.707\,\,\hbox{m/s}
	$$

{\it Example 3.} At $y_*=8.53\,\,$m,  $t_*=28.8\,\,$s, $H_*=0.7\,$cm, 
$g=9.8\,\,\hbox{m/s}^2$,  $2y_*/(gt_*^2)=0.0021$ we obtain  
$$
\tau=100,\,\,\,\,\tan\theta_f(100)=20.990,\,\,\,\,\tan\theta_r(100)=20.450,\,\,\,\,\tan\theta_f(100.1)=21.032
	$$
$$
h(100)=0.072,\,\,\,\,L_*=1.75 \,\,\hbox{cm},\,\,\,\,a=|f|L_*=0.406\,\,\hbox{m}
	$$
$$
h_*=1.3 \,\,\hbox{mm},\,\,\,\,\,\,\,\,l_*=0.22\,\,\hbox{m},\,\,\,\,
V_*=0.296\,\,\hbox{m/s},\,\,\,\,v_*=0.592\,\,\hbox{m/s}
	$$
The section $170<y<250$ in figure 4 (bottom panel, the curve on the left) of 
the packet includes only waves with amplitudes greater than 0.2 mm. The length 
of the packet is estimated as $(25-17)\cdot a=3.2\,$ m.                                                      

The examples show that the average speed of the packet of finite amplitude 
during the travel time $t_*$ is a half of the speed of the WMH at the instant 
$t_*$; exactly the same relationship exists between group and phase velocities 
in linear theory of deep water waves of infinitesimal amplitude; 
the average speed of the packet (8.2) and the speed of the WMH 
are equal, respectively, to the group and phase velocities of a linear wave packet provided that sinusoidal carrier wave of the linear packet and the wave of maximum height of nonlinear packet (8.2) have the same wavelength. 

The reason for the relationship between the four speeds 
is that (by their definitions) the speeds are independent of the amplitude 
of packet (8.2).

The assertion 2 and the examples lead to 

{\bf Assertion 3}:

The average speed of a nonlinear wave packet of finite amplitude during a travel time 
$t_*$ is a half of the speed of the packet's wave of maximum height at the instant $t_*$; 
the average speed of the nonlinear packet and the speed of the  wave of maximum height 
are equal, respectively, to the group and phase velocities of a linear wave packet  provided that sinusoidal carrier wave of the linear packet and the wave of maximum height of the nonlinear packet have the same wavelength. 

Now we return to the pulse experiment by Feir.
In [9], the length of the wave pulse shown in figure 1, the travel time $t_*$ 
from the wavemaker to the gauge and the maximum amplitude $H_*$ of the pulse 
are not shown. Nevertheless, the first two quantities can be estimated 
as follows:
by the linear theory, a wave packet with carrier frequency of 2.5 Hz 
travels at the group velocity of $0.31\,\,\hbox{m/s}$. This suggests that 
the length of the wave pulse is $0.31\,\,\hbox{m/s}\cdot 10\,\,\hbox{s}=3.1\,\,\hbox{m}$, 
and the travel time from the wavemaker to the gauge is about 27.5 s. 

These values of the travel time and length of the experimental 
wave pulse are in good agreement with 
the estimates obtained in the three examples above.
\bigskip

{\bf 11. About higher-order approximations to the full nonlinear problem.}
\bigskip

Under what conditions the solution of the leading-order equations gives a reasonable
approximation of solution for full nonlinear problem? 

The usual technical work with expansions leads to a sequence of systems
of linear nonhomogeneous equations. Every nonhomogeneous system of the
sequence involves terms corresponding to applied "forces" or "sources".
The corresponding homogeneous equations are identical. It is impossible
to answer the question unless the exact solution of the leading-order
equations is studied for arbitrary initial conditions and arbitrary "force"\,-terms. 

It can be shown that under boundary conditions (3.5) the "force"\,-terms have
the form $\cos^2\theta\cdot F(\theta, t)$. 

Having expanded the function $F(\theta, t)$ in a trigonometric series of the form
$$
F(\theta,t)=\sum_{k=1}^{+\infty}\alpha_k(t)\cos(2k\theta)+
\beta_k(t)\sin(2k\theta)
	$$
$$
\sum_{k=1}^{+\infty}\sqrt{{\alpha_k}^2+{\beta_k}^2}<+\infty.
    $$
we can obtain nonhomogeneous linear integrodifferential equations 
just of the type considered in [5] where the exact solution to the nonhomogeneous 
equations were obtained. 

The solution shows that the "input" and the "output" of the nonhomogeneous linear 
equations (i.e., the applied "forces" and the solution) are related by a bounded 
operator. 

This means that the exact solution 
of the leading-order equations give a reasonable approximation to the solution of the
full nonlinear problem for small values of the parameter $\varepsilon$. 

We are able to construct 
higher-order approximaations to the solution of the full nonlinear problem, 
provided the "force"\,-terms are actually expanded in the trigonometric series similar to that shown above.
\bigskip

{\bf Appendix A}

Solution to the leading-order equations (4.1), (4.2) is sought by expanding unknown functions in trigonometric series: 
$$
W_0(\theta,t)=\cos^2\theta\sum_{k=1}^{+\infty}
[\hat a_k(t)\cos (2k\theta)+\hat b_k(t)\sin (2k\theta)]  
        $$
                          $$\eqno(A1) $$   
$$
\nu_0(\theta,t)=\sum_{k=0}^{+\infty}
[\hat \rho_k(t)\cos (2k\theta)+\hat e_k(t)\sin (2k\theta)] 
        $$
In order to simplify matters, initial values of the coefficients are  assigned as
$$
\hat a_n(0)=a_n,\,\,\,\,\,\hat a_k(0)=0\,\,\,\hbox{for}\,\,\,k\ne n
\,\,\,\,\,\,\hat b_k(0)=\hat\rho_k(0)=\hat e_k(0)=0     \eqno(A2)    
        $$
Inserting (A.1) in the equations (4.1), (4.2),   
making use of the expressions (4.3) for operator $H(F)$, and 
equating coefficients of  $\cos(2k\theta)$, 
we obtain the following differential equations for coefficients of the series (A.1):
$$
\frac{d\hat a_k}{dt}=-\frac{k}{|f|}\hat\rho_k,\,\,\,\,\,\,
\frac{d\hat\rho_k}{dt}=\frac{1}{2}\hat a_{k-1}+\hat a_k+
\frac{1}{2}\hat a_{k+1},\,\,\,\,\,\,k=0,\,1,\,2,\,\dots        \eqno(A3)      
        $$
Equations (A.3) are derived using formulas (4.3) at small values of $|\sigma|$. Letting $\sigma$ tend to 0, we find one-sided limits shown in (4.1) and (4.2).

Exact solution to the initial value problem (A.3), (A.2) is given by 
$$
n\hat a_k(t)=(-1)^k(-1)^n2a_n \int\limits_0^{+\infty}
x^3e^{-x^2}L_{k-1}^{(1)}(x^2)L_{n-1}^{(1)}(x^2)\cos(\tau x)\,dx
	$$
                                                     $$\eqno(A.4)$$
$$
\frac{k}{|f|}\hat\rho_k=(-1)^k(-1)^n2a_n\int\limits_0^{+\infty}
x^4e^{-x^2}L_{k-1}^{(1)}(x^2)L_{n-1}^{(1)}(x^2)\sin(\tau x)\,dx
	$$
This can be verified by substituting (A.4) into equations (A.3) 
(formula (5.8) and the orthogonality integral for Laguerre polynomials should be taken 
into account).

Equations (A.1), (A.4) lead to expressions (5.3), (5.5) for $C_n(\tau,\theta)$ 
and $H_n(\tau,\theta)$. 
Expressions (5.4) and (5.6) for $S_n(\tau,\theta)$ 
and $G_n(\tau,\theta)$ can be obtained in the similar way.
\bigskip 

{\bf REFERENCES}
\bigskip

1. Zakharov, V.E.:  Stability of periodic waves of finite amplitude 
on the surface of a deep fluid,  J. App. Mech. Tech. Phys.  9, 190-194 (1968). 
(Translated from Russian)

2. Hasimoto, H., Ono, H.: Nonlinear modulation of gravity waves, 
 J. Phys. Soc. Jap. 33,  805-811 (1972).
                                                          
3. Davey, A.: The propagation of a weak nonlinear wave, 
 J. Fluid Mech. 53, 769-781 (1972).

4. Yuen, H.C., Lake, B.M.: Nonlinear Deep Water Waves: Theory 
and Ex\-pe\-ri\-ment, Phys. Fluids.  18, 956-960 (1975).

5. Zakharov, V.E., Shabat, A.B.: Exact theory of two-dimensional 
self-focusing and one-dimensional self-modulating waves in nonlinear media, 
 Soviet Phys. JETF. 65,  997-1011 (1972). (Translated from Russian)

6. Mindlin, I.M.: Nonlinear waves in two-dimensions generated by variable 
pressure acting on the free surface of a heavy liquid, 
 J. Appl. Math. Phys. (ZAMP) 55, 781-799 (2004).

7. Mindlin, I.M.: Integrodifferential Equations in Dynamics of Heavy     
Layered Liquid,  Nauka$\ast$Fizmatlit, Moscow,(1996) (Russian).

8. Yuen, H.C., Lake, B.M.:  Nonlinear dynamics of deep-water 
gravity waves. In:  Chia-Shun Yih (Ed.), Advances in applied mechanics, 22,
Academic press, New York  $\ast$ London, 67-229 (1982).

9. Feir, J.E.:  Discussion: Some results from wave pulse experiments, 
In: A discussion on nonlinear theory of wave propagation 
in dispersive systems,  Proc. R. Soc.  299 London, 54-58 (1967).

10. Show, R.P.: Boundary integral equation methods applied to water waves. In: T.A.Cruze, F.J. Rizzo (Ed.), Boundary-integral equation method: computational applications in applied mechanics, ASME, Rensselaer Polytechnic Institute, June 23-25,. AMD-vol. 11 (1975)

11. Suetin, P.K.  Classical Orthogonal Polynomials,  
Nauka, Moscow,  (1964) (Russian).

\end{document}